\documentclass[12pt]{article}%
\usepackage{amsmath}
\usepackage{amssymb}%
\usepackage{amsfonts}%
\usepackage{graphicx}
\input epsf.tex
\newcommand{\be}{\begin{equation}}
\newcommand{\ee}{\end{equation}}
\newcommand{\bea}{\begin{eqnarray}}
\newcommand{\eea}{\end{eqnarray}}

\begin{document}
\bigskip\begin{titlepage}
\begin{flushright}
UUITP-11/10\\
\end{flushright}
\vspace{1cm}
\begin{center}
{\Large\bf Entropic dark energy and sourced Friedmann equations\\}
\end{center}
\vspace{3mm}
\begin{center}
{\large
Ulf H.\ Danielsson} \\
\vspace{5mm}
Institutionen f\"or fysik och astronomi, Box 516, SE-751 20
Uppsala, Sweden
\vspace{5mm}
{\tt
ulf.danielsson@physics.uu.se \\
}
\end{center}
\vspace{5mm}
\begin{center}
{\large \bf Abstract}
\end{center}
\noindent
In this paper we show that a recent attempt to derive dark energy as an entropic force suffers from the same problems
as earlier attempts motivated by holography. A possible  remedy is again the introduction of source terms.
\vfill
\begin{flushleft}
March 2010
\end{flushleft}
\end{titlepage}\newpage


\section{Introduction}

The discovery by Bekenstein, \ \cite{Bekenstein:ax}, that black hole physics
lead to thermodynamics, have by many been seen as an important clue to the
nature of gravity and its quantization. An especially intriguing step forward
in this context was Jacobson's realization, \cite{Jacobson:1995ab}, that the
argument also goes in the other direction: you can start with thermodynamics
and derive the Einstein equations from there. Many have tried to extend these
ideas using holography, and apply them to, among other things, cosmology. The
progress has been meager.

In particular, there has been attempts to derive dark energy using this kind
of reasoning. Some examples can be found in \cite{Cohen:1998zx} and
\cite{Horvat:2004vn}. The shortest line of reasoning goes like this. The
number of degrees of freedom in a Hubble volume is limited by holography to be
given by the area of the horizon. That is, $R^{2}$ in planckian units where
$R\sim1/H$. To each of these degrees of freedom you may associate a zero point
energy of the order $1/R$. The total energy is then $\sim R^{2}\times1/R\sim
R$. Spreading this out over the volume $R^{3}$ gives an energy density
$1/R^{2}\sim H^{2}$. Hence this is of the same order as the cosmological constant.

This approach suffers from many problems when you look at the details. The
most obvious one is that the mere addition of a component with energy given by
$H^{2},$ leads to the wrong equation of state: the universe will not
accelerate. As we will review below, the situation can be improved if we
introduce source terms, \cite{Horvat:2004vn},\cite{Danielsson:2004xw}.

In \cite{Verlinde:2010hp} an attempt was made to further develope the ideas of
Jacobson. Whether this succeeds or not, depends on whether new conclusions can
be reached concerning, e.g., cosmology. This is the subject of a recent paper,
\cite{Easson:2010av}, where it is claimed that the results of
\cite{Verlinde:2010hp} automatically leads to an explanation of dark energy.
In this note I will put this claim in the context of previous research, and
argue that the status of the subject is unchanged.

\section{Modifying the Friedmann equations}

\bigskip

In \cite{Easson:2010av} it was proposed that one of the Friedmann equations
should be corrected by an extra term, $s$, according to%
\begin{equation}
\frac{\ddot{a}}{a}=-\frac{4\pi}{3M_{p}^{2}}\left(  \rho+3p\right)
+s.\label{surfmod}%
\end{equation}
It was argued that such a term can be motiviated from the possible presence of
usually discarded boundary terms in the Einstein equations. Alternatively, it
can be argued to arise due to the presence of the entropic force discussed in
\cite{Verlinde:2010hp}. If this argument is correct, does this mean that
physicists over the years have missed an obvious term in the Friedmann
equations? Is the claim in \cite{Verlinde:2010hp} that the entropic
formulation is equivalent to ordinary gravity, or does it give new and
different predicitons in the context of cosmology? What I want to argue for
below is that a modification of this kind does not represent something
fundamentally different compared to what has already been considered.

Since we want to approach gravity and cosmology from a thermodynamical point
of view, we should first make sure that the equations we write down are
manifestly in this spirit. Following \cite{Danielsson:2004xw} and
\cite{Frolov:2002va}, one can apply the reasoning of Jacobson in a simplified
cosmological setting. The starting point is the relation between the area of
the horizon, and entropy for a black hole given by%
\begin{equation}
S=\frac{M_{p}^{2}}{4}A.
\end{equation}
We assume, like in \cite{GibbHawk:1977}, that the cosmological horizon
determined by the Hubble constant plays a similar role as the horizon of a
black hole, and assign an entropy to the horizon according to%
\begin{equation}
S=\frac{\pi M_{p}^{2}}{H^{2}}.
\end{equation}
From this we can now derive the Friedmann equations using the standard
relation between flow of heat and entropy, $dQ=TdS$. The flow of heat out
through the horizon is then related to a flow of entropy given by
\begin{equation}
\dot{Q}=\dot{S}T=A\left(  \rho+p\right)  ,
\end{equation}
where
\begin{equation}
T=\frac{H}{2\pi}.
\end{equation}
Expressing the entropy in terms of the horizon area and the Hubble constant,
we find%
\begin{equation}
\dot{H}=-\frac{4\pi}{M_{p}^{2}}\left(  \rho+p\right)  ,\label{eq:h0prick}%
\end{equation}
which, indeed, is one of the Friedmann equations.

To completely specify the time evolution we also need the continuity equation%
\begin{equation}
\overset{\cdot}{\rho}+3H\left(  \rho+p\right)  =0,\label{conteq}%
\end{equation}
which, combined with (\ref{eq:h0prick}), give rise to another of the Friedmann
equations, i.e.,%

\begin{equation}
H^{2}=\frac{8\pi}{3M_{p}^{2}}\rho+\mathrm{const.}\label{eq:h02}%
\end{equation}
Usually, the two Friedmann equations together with the continuum equation are
viewed on an equal footing keeping in mind that only two of them are
independent. However, from our thermodynamical point of view, there is an
important difference between the various choices. (\ref{eq:h02}) is obtained
from (\ref{eq:h0prick}) using integration and there is, therefore, a
corresponding constant of integration, \textit{the cosmological constant},
which does not appear in the basic equations (\ref{eq:h0prick}). The usual
interpretation is that the cosmological constant corresponds to matter with
$p=-\rho$. However, from our thermodynamical point of view, it might be more
natural to view the cosmological constant as part of the initial conditions.

Let us now come back to the modifed equation (\ref{surfmod}). It is easy to
see that unless $s$ is a constant, (\ref{surfmod}) is inconsistent with
(\ref{eq:h0prick}) and (\ref{conteq}). (\ref{eq:h0prick}) lies at the heart of
thermodynamics and holography and should not be modified.\footnote{In
principle one can also modify equation (\ref{eq:h0prick}) \ demanding instead
that there are no source terms for matter. Since, according to
\cite{Jacobson:1995ab}, there is also a local version of (\ref{eq:h0prick}),
i.e. the Einstein equations, such a procedure implies that\ the entropic dark
energy turns into a local fluid with very specific properties. It is not clear
how to interprete this from a holographic point of view, and this option will
not be considered further in this paper.} (\ref{conteq}), on the other hand,
can be modified in a simple way by adding a source term:%
\begin{equation}
\overset{\cdot}{\rho}+3H\left(  \rho+p\right)  =q,
\end{equation}
which is easily seen to lead to%
\begin{equation}
H^{2}=\frac{8\pi}{3M_{p}^{2}}\rho-\frac{8\pi}{3M_{p}^{2}}\int^{t}qdt.
\end{equation}
Furthermore, we find that the source term is related to the extra term in
(\ref{conteq}) through%
\begin{equation}
q=-\frac{3M_{p}^{2}}{8\pi}\dot{s}.\label{eq:qsprick}%
\end{equation}
One can also, directly from (\ref{surfmod}) using (\ref{eq:h0prick}), arrive
at%
\begin{equation}
H^{2}=\frac{8\pi}{3M_{p}^{2}}\rho+s.\label{eq:h2rs}%
\end{equation}

We conclude that that the correcting terms discussed in \cite{Easson:2010av},
which are not constant in time, necessarily imply the inclusion of a source
term for matter if we remain true to the thermodynamic view on gravity. The
interpretation is that the would be cosmological constant, arising as a
constant of integration, is promoted into a running dark energy. It is
depleted as time goes by and the energy is converted into ordinary matter or
radiation through the source term.

It is important to realize that the naive application of (\ref{surfmod}),
assuming $\rho\sim a^{-3\left(  1+w\right)  }$, will not lead to the correct
results in this setup.\footnote{See, however, previous footnote.} The form of
$\rho$ will be different due to the energy creation, and a more careful
analysis is necessary. This is what we turn to next.

\section{Solving the sourced Friedmann equations}

To proceed we need to be more specific. Let us assume that there are one
matter component that is sourced by the dark energy and, for generality, one
that is not. We have%

\begin{align}
\overset{\cdot}{\rho}_{s}+3H\left(  \rho_{s}+p_{s}\right)    & =q\\
\overset{\cdot}{\rho}_{us}+3H\left(  \rho_{us}+p_{us}\right)    & =0.
\end{align}
We assume that the equations of state are such that $p_{s}=w_{s}\rho_{s}$ and
$p_{us}=w_{us}\rho_{us}$. One can then show, slightly generalizing
\cite{Danielsson:2004xw}, that%
\begin{equation}
a^{2}HH^{\prime\prime}+a^{2}H^{\prime2}+\left(  4+3w_{s}\right)  aHH^{\prime
}=-\frac{4\pi}{M_{p}^{2}}\left(  1+w_{s}\right)  H^{-1}q+\frac{12\pi}%
{M_{p}^{2}}\left(  1+w_{us}\right)  \left(  w_{us}-w_{s}\right)  \rho_{us}.
\end{equation}
This is a second order differential equation for $H$, yielding two constants
of integration. One will generalize the cosmological constant, and the other
one will fix $\rho_{s}$ at some moment in time.

Below we will focus on the case where the sourced matter is in the form of
radiation. We will consider two examples of sources.

\subsection{Entropic dark energy}

\bigskip

To capture the entropic approach to dark energy we consider $s=kH^{2}$ for
some numerical constant $k$. The value of $k$ does not affect the conclusions
in any essential way. We find%

\begin{equation}
a^{2}HH^{\prime\prime}+a^{2}H^{\prime2}+5aHH^{\prime}=4kaHH^{\prime}%
+\frac{12\pi}{M_{p}^{2}}\left(  1+w_{us}\right)  \left(  w_{us}-w_{s}\right)
\rho_{us}.
\end{equation}
The homogenous solution is%
\[
H^{2}=C_{1}+C_{2}a^{-\frac{4}{1+4k}},
\]
where we note the presence of a cosmological constant, and a would be
radiation component that is decaying less rapidly than $a^{-4}$ (if $k>0$).
What happens is that the dark energy proportional to $H^{2}$ decays as $H$
decreases, and the energy is converted into radiation.

Going back to (\ref{eq:h2rs}), we see that the cosmological constant
necessarily is part of $\rho$ and $p$. Such a component is necessary in order
for $H^{2}$ to be close to constant. Ironically, we conclude that the
cosmological constant is a free parameter, and manifestly \textit{not} given
by the entropic contribution in this setup.

\subsection{An example related to transplanckian physics}

An extra source term can also be connected with a modified vacuum. This has
been done in \cite{Danielsson:2004xw}. There it was assumed that some of the
matter or radiation fields were put in an excited state at small scales and
high energy. In the case of inflation this would correspond to a deviation
from the usual Bunch-Davies vacuum. This requires the introduction of a source
term that injects the necessary energy into the system. The amount of modes in
this excited state is given by the Bogolubov coefficient $\left\vert \beta
_{k}\right\vert ^{2}$. The total energy generated in this way is given by%

\begin{equation}
\rho_{\Lambda}\left(  a\right)  =\frac{3}{2\pi}\int_{\varepsilon}^{\Lambda
}dpp^{3}\left\vert \beta_{\frac{ap}{\Lambda}}\right\vert ^{2}=\frac{3}{2\pi
}\frac{\Lambda^{4}}{a^{4}}\int_{a_{i}}^{a}dxx^{3}\left\vert \beta
_{x}\right\vert ^{2},
\end{equation}
where we have introduced a low energy cutoff corresponding to the present
energy of modes that started out at $\Lambda$ at a time when the Hubble
constant was as small as $H_{i}$.

A specific case of vacuum fluctuations, suggested in \cite{Danielsson:2002kx},
with the characteristic values of the Bogolubov mixing given by%
\begin{equation}
\left\vert \beta_{k}\right\vert ^{2}\sim\frac{H^{2}}{\Lambda^{2}},
\end{equation}
was studied in \cite{Danielsson:2004xw}. If we take a derivative of the energy
density with respect to the scale factor and use $\frac{d}{da}=\frac{1}%
{aH}\frac{d}{dt}$, we find%
\begin{equation}
\dot{\rho}_{s}+4H\rho_{s}=\frac{3}{2\pi}\Lambda^{2}H^{3},\label{eq:contQ}%
\end{equation}
and we can conclude that the source term is given by%
\begin{equation}
q=\frac{3}{2\pi}\Lambda^{2}H^{3}.
\end{equation}

If $\Lambda\ll M_{p}$ then we find (ignoring $\rho_{us}$)
\begin{equation}
H^{2}=C_{1}a^{-2n_{1}}+C_{2}a^{-2n_{2}}\label{eq:h2c1c2}%
\end{equation}
where $C_{1,2}$ are constants of integration, and%
\begin{equation}
n_{1,2}=1\pm\sqrt{1-\frac{4\Lambda^{2}}{M_{p}^{2}}}.
\end{equation}
That is, one term corresponding to a slowly decaying dark energy, and term
corresponding to radiation decaying more slowly than usual due to energy transfer.

For these kind of sources we find that \textit{ }one can not assign an
unambiguous value to the cosmological constant.\textit{ }We find, instead,
that a fixed dimensionful cosmological constant, is effectively replaced by a
dimensionless parameter determining the running, given by the ratio of a
fundamental scale and the Planck scale.

\section{Discussion}

\bigskip

We have seen that the modified Friedmann equation (\ref{surfmod}) of
\cite{Easson:2010av} under some reasonable assumptions is equivalent to the
addition of a source term to the continuity equation. The explicit example
considered in \cite{Easson:2010av}, motivated by the entropic force, has the
Friedmann equation modified by $s\sim H^{2}.$ This is very similar to previous
attempts to argue for the presence of a holographic energy of order $H^{2}$.
As shown already in \cite{Horvat:2004vn} and \cite{Danielsson:2004xw}, this
can be accomodated through the addition of a source term obtainable through
(\ref{eq:qsprick}). The conclusion is that the results of \cite{Easson:2010av}
do not go beyond those obtained previously in the literature. The motivation
for how the dark energy component arises is subtly different but in essence
the same.

The relation between gravity and thermodynamics and the possible consequences
for cosmology and dark energy remain as elusive as ever.

\section*{Acknowledgments}

The work is supported by the Swedish Research Council (VR) and the G\"{o}ran
Gustafsson Foundation.

\end{document}